\documentclass[aoas,MSNbibl,nameyear,seceqn,dvips]{arximspdf}
\usepackage{dcolumn}
\usepackage{graphicx}

\doi{10.1214/12-AOAS557} 
\volume{6}
\issue{4}
\pubyear{2012}
\firstpage{1689}
\lastpage{1706}

\makeatletter

\newcolumntype{d}[1]{D{.}{.}{#1}}

\newcommand{\eqref}[1]{(\ref{#1})}

\newcommand{\iint}{\int\!\!\int}

\makeatother

\begin{document}
\begin{frontmatter}

\title{Truth and memory: Linking instantaneous and retrospective
self-reported cigarette consumption}
\runtitle{Self-reported cigarette consumption}

\begin{aug}
\author[A]{\fnms{Hao} \snm{Wang}\corref{}\thanksref{t1,t2}\ead[label=e1]{hwang76@jhmi.edu}},
\author[B]{\fnms{Saul} \snm{Shiffman}\ead[label=e2]{shiffman@pinneyassociates.com}},
\author[C]{\fnms{Sandra D.} \snm{Griffith}\thanksref{t1}\ead[label=e3]{sgrif@mail.med.upenn.edu}}\\
\and
\author[D]{\fnms{Daniel F.} \snm{Heitjan}\thanksref{t3}\ead[label=e4]{dheitjan@upenn.edu}}
\runauthor{Wang, Shiffman, Griffith and Heitjan}
\affiliation{Johns Hopkins University,
University of Pittsburgh, University~of~Pennsylvania and
University~of~Pennsylvania}
\address[A]{H. Wang\\
Division of Biostatistics and Bioinformatics\\
Sidney Kimmel Comprehensive Cancer Center\\
Johns Hopkins Medicine\\
550 N. Broadway, Suite 1103\\
Baltimore, Maryland 21218\\
USA\\
\printead{e1}}
\address[B]{S. Shiffman\\
Department of Psychology\\
University of Pittsburgh\\
BELPB 510\\
Pittsburgh, Pennsylvania 15260\\
USA\\
\printead{e2}}
\address[C]{S. D. Griffith\\
Department of Biostatistics\\
\quad and Epidemiology\\
University of Pennsylvania\\
503 Blockley Hall\\
423 Guardian Drive\\
Philadelphia, Pennsylvania 19104\\
USA\\
\printead{e3}}
\address[D]{D. F. Heitjan\\
Department of Biostatistics\\
\quad and Epidemiology\\
University of Pennsylvania\\
622 Blockley Hall\\
423 Guardian Drive\\
Philadelphia, Pennsylvania 19104\hspace*{2pt}\\
USA\\
\printead{e4}} 
\end{aug}

\thankstext{t1}{Supported by USPHS Grant T32-CA093283.} 
\thankstext{t2}{Supported by Maryland Cigarette Restitution Fund
Research grant to the Johns Hopkins Medical Institutions.} 
\thankstext{t3}{Supported by USPHS Grant CA R01-CA116723.}

\received{\smonth{7} \syear{2010}}
\revised{\smonth{2} \syear{2012}}

%
\begin{abstract}
Studies of smoking behavior commonly use the \textit{time-line
follow-back} (\textit{TLFB}) method, or periodic retrospective recall,
to gather data on daily cigarette consumption. TLFB is considered
adequate for identifying periods of abstinence and lapse but not for
measurement of daily cigarette consumption, thanks to substantial
recall and digit preference biases. With the development of the
hand-held electronic diary (ED), it has become possible to collect
cigarette consumption data using \textit{ecological momentary
assessment} (\textit{EMA}), or the instantaneous recording of each
cigarette as it is smoked. EMA data, because they do not rely on
retrospective recall, are thought to more accurately measure cigarette
consumption. In this article we present an analysis of consumption data
collected simultaneously by both methods from 236 active smokers in the
pre-quit phase of a smoking cessation study. We define a statistical
model that describes the genesis of the TLFB records as a two-stage
process of mis-remembering and rounding, including fixed and random
effects at each stage. We use Bayesian methods to estimate the model,
and we evaluate its adequacy by studying histograms of imputed values
of the latent remembered cigarette count. Our analysis suggests that
both mis-remembering and heaping contribute substantially to the
distortion of self-reported cigarette counts. Higher nicotine
dependence, white ethnicity and male sex are associated with greater
remembered smoking given the EMA count. The model is potentially useful
in other applications where it is desirable to understand the process
by which subjects remember and report true observations.
\end{abstract}

%

\end{frontmatter}

\section{Introduction}\label{sec1}\label{chp3intro}
A common technique for eliciting consumption in studies of substance
abuse is the time-line follow-back (TLFB) method, in which one asks
subjects to report daily consumption retrospectively over the preceding
week, month or other designated period. In smoking cessation research,
for example, TLFB is one important method for measuring cigarette
consumption and defining periods of quit and lapse.

Although TLFB is a practical approach to quantifying average smoking
behavior [\citet{Brown}], TLFB data can harbor substantial
errors as measures of daily consumption [\citet{Klesges}].
TLFB questionnaires request exact daily cigarette counts, which smokers
are unlikely to remember, particularly after several days have passed.
Moreover, some smokers may understate consumption to avoid the social
stigma attached to excessive smoking or an inability to quit
[\citet{Boyd}]. Thus, smoking cessation studies typically require
validation of TLFB reports of zero consumption by biochemical
measurement of exhaled carbon monoxide or nicotine metabolites from
saliva or blood.

A second concern is that histograms of TLFB-derived daily cigarette
counts commonly exhibit spikes at multiples of 20, 10 or even 5
cigarettes. This phenomenon, known as ``digit preference'' or
``heaping,'' is thought to reflect a tendency to report consumption in
terms of packs (each pack in the US contains 20 cigarettes) or half or
quarter packs. The heaps presumably arise because many smokers do not
remember precisely how many cigarettes they smoked and therefore report
their count rounded off to a nearby convenient number. It has also been
hypothesized that some smokers consume exactly an integral number of
packs per day as a self-rationing strategy [\citet{Farrell}], but
evidence so far suggests that such behavior, if it exists, causes only
a small fraction of the observed heaping [\citet{WangHeitjan}].
Indeed, \citet{Klesges} observed that the distribution of
biochemical residues of smoking is smooth, suggesting that heaping is a
phenomenon of reporting rather than consumption.

Recall bias and heaping bias in self-reported longitudinal cigarette
counts potentially affect estimates of both means and treatment
effects. Moreover, heaping may lead to underestimation of
within-subject variability, thanks to smokers who regularly report one
pack rather than a precise count that varies around some mean in the
vicinity of 20. If a large enough fraction of subjects in a study are
of this kind, estimates of both within-subject and between-subject
variability can be distorted.

Although there has been substantial research on statistical modeling of
heaping and digit preference in a range of disciplines
[Heitjan and Rubin (\citeyear{HeitjanRubin,HR2}),
\citet{RidoutMorgan}, \citet{Pickering},
\citet{Klerman}, \citet{Torelli}, \citet{Dellaportas},
\citet{Roberts}, \citet{Wright} and \citet{Wolff}], the
only such application in smoking cessation research is that of
\citet{WangHeitjan}, who described a latent-variable rounding
model for heaped univariate TLFB cigarette count data. They postulated
that the reported cigarette count is a function of the unobserved true
count and a latent heaping behavior variable. The latter can take one
of four values, representing exact reporting, rounding to the nearest
5, rounding to the nearest 10, and rounding to the nearest 20. Except
for ``exact'' reporters (i.e., those who report counts not divisible by
5), one obtains at best partial information on the true count and the
heaping behavior. They analyzed univariate count data from a smoking
cessation clinical trial, assuming a zero-inflated negative binomial
distribution for the true underlying counts together with an ordered
categorical logistic selection model for heaping behavior given true
count.

The analysis of \citet{WangHeitjan} has three important
limitations: first, they included only data from the last day of eight
weeks of treatment, ignoring the 55 preceding days. Second, they
assumed---without empirical verification---that reported counts not
divisible by 5 were accurate. And third, they assumed that the
preference for counts ending in 0 or 5 actually represented rounding
rather than some other form of reporting error. That is, a declared
count of 20 cigarettes was taken to mean that the true count was
somewhere between 10 and 30 cigarettes, and was merely misreported as
20. In the absence of more accurate data on the true, underlying count,
attempts to model heaping must rely on some such assumptions.

Precise assessment of smoking behavior has taken on increasing
importance as researchers explore the value of reducing consumption as
a way to lessen the harms of smoking [\citet{shiffman1},
\citet{Hatsukami}] and to improve the chance of ultimately
quitting [\citet{Shiffman3}, \citet{Cheong}]. The advent of
the inexpensive hand-held electronic diary (ED) that allows the
instantaneous recording of \textit{ad libitum} smoking has created the
possibility of making much more accurate measurements. Such evaluation
is an instance of \textit{ecological momentary assessment} [EMA;
\citet{StoneShiffman}], in that it generates records of events
logged as they occur in real-life settings. In \citet{Shiffman2},
researchers asked 236 participants in a smoking cessation study to use
a specially programmed ED to record each cigarette as it was smoked
over a 16-day pre-quit period; moreover, the ED periodically prompted
the smokers to record any cigarettes they had missed. At days 3, 8 and
15, subjects visited the clinic to complete a TLFB assessment of daily
smoking since the preceding visit (2, 5 or 7 days previously), stating
how many cigarettes they had smoked each day. The study found that
while the TLFB data contained the expected heaps at multiples of 10 and
20, the EMA data had practically none. Average smoking rates from the
two methods were moderately correlated ($r=0.77$), but the
within-subject correlation of daily consumption between TLFB and EMA
was modest ($r=0.29$). Self-report TLFB consumption was on average
higher than EMA (by 2.5 cigarettes), but on 32\% of days, subjects
recorded more cigarettes by EMA than they later recalled by TLFB.

These data provide us with an opportunity---unprecedented, so far as we
know---to study the relationship between self-reports of daily
cigarette consumption by TLFB and EMA. To describe this relationship,
we develop a statistical model with two components: the first is a
regression that predicts the patient's notional ``remembered''
cigarette count (a latent factor) from the EMA count. The second is a
regression that predicts the rounding behavior---described as in
\citet{WangHeitjan} with an ordinal logistic regression---from the
remembered count and fully observed predictors. The models include
random subject effects that describe the propensities of the subjects
to mis-remember their actual consumption (in the first component) and
to report the remembered consumption with a characteristic degree of
accuracy (in the second). Assuming that EMA represents the true count,
the first component of the model allows us to examine the recall bias
resulting from mis-remembering, while the second component describes
the heaped reporting errors.

\section{Notation and model}\label{sec2}\label{method1}
Let $Y_{it}$ denote the observed heaped TLFB consumption for subject
$i$ on day $t$, $i=1, \ldots, n$, $t=1, \ldots, m_i$, and let
$Y_i=(Y_{i1}, \ldots, Y_{im_i})^{T}$ denote the vector of TLFB data
for subject $i$. Let $X_{it}$ be the EMA consumption on subject $i$,
day $t$, and let $X_i=(X_{i1}, \ldots, X_{im_i})^{T}$ be the vector of
EMA data for subject $i$. We furthermore let $Z_i=(Z_i^R,Z_i^H)$ be a
vector of baseline predictors for subject $i$, with $Z_i^R$
representing predictors of recall and $Z_i^H$ predictors of heaping.
These predictor sets may overlap.

\subsection{A model for remembered cigarette count}\label{sec2.1}
The first part of our model assumes that for each day and subject there
is a notional remembered cigarette count, denoted $W_{it}$ [$W_i=
(W_{i1}, \ldots,W_{im_i})^{T}$]. We assume $W_{it}$ is distributed as
Poisson conditionally on a random effect $b_i$, the EMA smoking pattern
$X_{it}$ and the covariate vector $Z_i$, with mean
%
%
\begin{equation}\label{recall} 
{\mathrm E}(W_{it}|X_{it},Z_i,b_i)= \exp\bigl(\beta_0+\ln(X_{it})\beta
_1+Z_i^R\beta_2+b_i\bigr).
\end{equation}
The parameters $\beta_1$ and $\beta_2$ represent the effects of EMA
consumption and baseline predictors, respectively, on the latent
remembered count. The random effect $b_i$, which we assume normally
distributed with mean $0$ and variance $\sigma_b^2$, represents
heterogeneity among subjects. We note that there are no $0$ values of
$X_{it}$ in the Shiffman data, which are from a pre-quit study in which
subjects were encouraged to smoke as normal. Thus, we can include $\ln
(X_{it})$ as a predictor. In more general contexts where $0$ EMA counts
are possible, one can adjust the model in simple ways to avoid this
problem. Moreover, when excessive $0$ counts occur in the TLFB data,
one can fit a zero-inflated count model, as in \citet{WangHeitjan},
for the remembered count.

\subsection{A model for the latent heaping process}\label{sec2.2}
Following \citet{WangHeitjan}, we assume that a latent rounding
indicator $G_{it}$ [$G_i=(G_{i1},\ldots, G_{im_i})^{T}$] dictates the
degree of rounding to be applied to the notional remembered
count~$W_{it}$. Specifically, we let $G_{it}$ take one of four possible
values: $G_{it}=1$ implies reporting the exact count, $G_{it}=2$
implies rounding to the nearest multiple of 5, $G_{it}=3$ implies
rounding to the nearest multiple of 10, and $G_{it}=4$ implies rounding
to the nearest multiple of 20. We assume that the probability
distribution of the heaping indicator depends on $W_{it}$, a
subject-level random effect $u_i \sim N(0,\sigma_u^2)$ that is
independent of $b_i$, and a baseline predictor vector $Z_i^H$.
Specifically, we propose the following proportional odds model for the
conditional distribution of~$G_{it}$:
%
%
\begin{equation}
\label{misreport} 
f(G_{it}|W_{it},Z_i,u_i) = \cases{
1-q(\gamma_1+\eta_{it}+u_i), &\quad if $g=1$;\cr
q(\gamma_1+\eta_{it}+u_i)-q(\gamma_2+\eta_{it}+u_i), &\quad
if $g=2$;\cr
q(\gamma_2+\eta_{it}+u_i)-
q(\gamma_3+\eta_{it}+u_i), &\quad if $g=3$;\cr
q(\gamma_3+\eta_{it}+u_i), &\quad if $g=4$.}\hspace*{-30pt}
\end{equation}
Here $\eta_{it}=W_{it}\gamma_0+Z_i^H\beta_3$, and $q(\cdot)$ is the
inverse logit function $q(x) = \exp(x)/\break(1+\exp(x))$. The parameters
$\gamma_1 > \gamma_2 > \gamma_3$ refer to the successive intercepts of
the logistic regressions, $\gamma_0$ refers to its slope with respect
to the remembered count, and $\beta_3$ refers to its slopes with
respect to the vector of heaping predictors $Z_i^H$. The random effect
$u_i$ describes between-subject differences in heaping propensity not
otherwise accounted for in the model.

\subsection{The coarsening function}\label{sec2.3}
\begin{figure}

\includegraphics{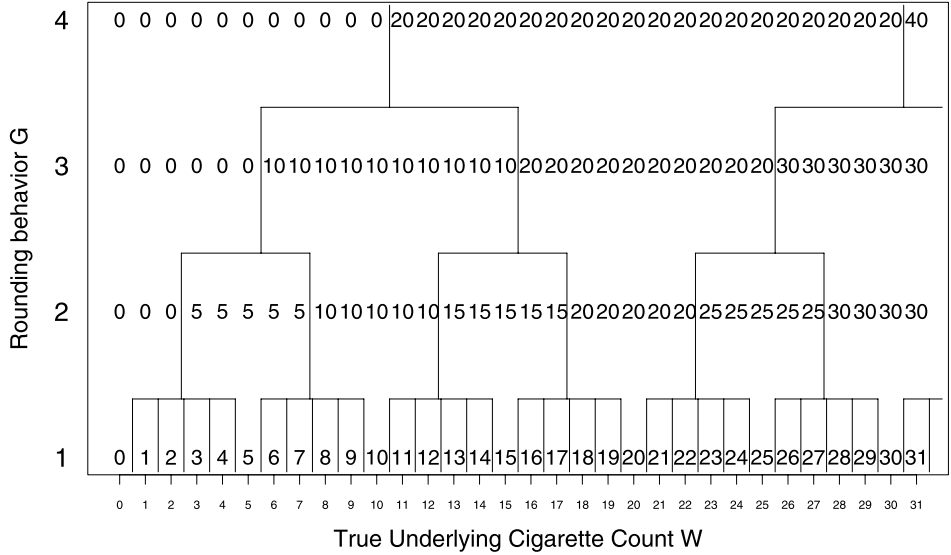}

\caption{Reported cigarette count $Y$ as a function
of the underlying count $W$ and the rounding behavior $G$.}\label{heapdiag3}
\end{figure}

As in \citet{WangHeitjan}, the model links the observed $Y_{it}$ to
the latent $W_{it}$ and $G_{it}$ via the coarsening function $h(\cdot
,\cdot)$:
\[
Y_{it} = h(W_{it},G_{it}),\qquad i=1,\ldots,n, t=1,\ldots,m_i.
\]
For example, at time $t$, subject $i$ with $W_{it}=14$ and $G_{it}=1$
reports $h(14,1)=14$, whereas $h(14,2)=15$, $h(14,3)=10$, and
$h(14,4)=20$. Figure~\ref{heapdiag3} illustrates this heaping mechanism.

%

A coarsened outcome $y_{it}$ may arise from possibly several
$(w_{it},g_{it})$ pairs. We denote the set of such pairs as
$\mathrm{WG}(y_{it}) = \{(w_{it},g_{it})\dvtx y_{it}=h(w_{it},g_{it})\}
$. For
example, a reported consumption of $y_{it}=5$ may represent a precise
unrounded value [$(w_{it}, g_{it})=(5,1)$] or rounding across a range
of nearby values [$(w_{it}, g_{it}) \in\{
(3,2),(4,2),(5,2),(6,2),(7,2)\}$]. For subject $i$, the probability of
the observed $y_{it}$ at time $t$ is the sum of the probabilities of
the $(w_{it}, g_{it})$ pairs that would give rise to it. The density of
reported consumption $y_{it}$ given the random effects can therefore be
expressed as
\[
f(y_{it}|b_i, u_i) = \sum_{(w_{it},g_{it})\in\mathrm{WG}(y_{it})}
f(w_{it}|b_i)f(g_{it}|w_{it},u_i).
\]

\subsection{Estimation}\label{sec2.4}
We estimate the model by a Bayesian approach that employs importance
sampling [\citet{Gelmancarlin}, \citet{Tanner}] to avoid iterative
simulation of parameters. The steps are as follows: we first compute
the posterior mode and information using a quasi-Newton method with
finite-difference derivatives [\citet{Dennis}]. We then
approximate the posterior with a multivariate $t_5$ density with mean
equal to the posterior mode and dispersion equal to the inverse of the
posterior information matrix at the mode. Next, we draw a large number
(4000) of samples from this proposal distribution, at each draw
computing the importance ratio $r$ of the true posterior density to the
proposal density. We then use sampling-importance resampling (SIR) to
improve the approximation of the posterior [\citet{Gelmancarlin}].
We evaluate posterior moments by averaging functions of the
simulated parameter draws with the importance ratios $r$ as weights.
The choice of a $t$ with a small number of degrees of freedom as the
importance density is intended to balance the convergence of the MC
integrals and the efficiency of the simulation.

Letting $\theta=(\beta_0,\beta_1,\beta_2,\beta_3,\sigma_b,\gamma
_1,\gamma_2,\gamma_3,\gamma_0,\sigma_u)$, the likelihood contribution
from subject $i$ is
%
%
\begin{eqnarray}\label{marginallik} 
L(\theta; y_i)&=&\iint\prod_{t=1}^{m_i}\sum_{(w_{it},g_{it})\in
\mathrm{WG}(y_{it})} f(w_{it}|b_i)f(g_{it}|w_{it}, u_i)\nonumber\\[-8pt]\\[-8pt]
&&\hspace*{100pt}{}\times f(b_i)f(u_i)\,
db_i \,du_i;\nonumber
\end{eqnarray}
we approximate the integral in \eqref{marginallik} by Gaussian
quadrature. We choose proper but vague priors for the parameters, which
we assume are a priori independent (except for $\gamma
_j, j=1,2,3$, as noted below). The parameter $\beta_1$ in the Poisson
mixed model~\eqref{recall}, representing the slope of the latent recall
on the EMA recorded consumption, is given a normal prior $\beta_1 \sim
N(1,10^2)$, whereas the priors of the other regression parameters in
both model parts are set to $N(0,10^2)$ subject to the constraint
$\gamma_1 > \gamma_2 > \gamma_3$. We assign the random-effect variances
inverse-gamma priors with mean and SD both equal to 1, a reasonably
vague specification [\citet{Carlin}]. We obtain the posterior
mode and information using SAS PROC NLMIXED, and implement Bayesian
importance sampling in \texttt{R}.

\section{Model checking}\label{sec3}\label{modelchecking}
With heaped data, the unavailability of simple graphical diagnostics
such as residual plots complicates model evaluation. We therefore
resort to examination of repeated draws of latent quantities from their
posterior distributions, in the spirit of Bayesian posterior predictive
checks [\citet{Rubin}, \citet{GelmanMeng}, \citet
{GelmanHeitjan}].
Specifically, we evaluate the adequacy of model
assumptions using imputed values of the latent recall $W$, which we
compare to its implied marginal distribution under the model.

Imputations of latent $W_i$ and $G_i$ are ultimately based on the
posterior density $f(\theta|y_i)$ of the model parameter $\theta$ given
the observed data $y_i$. \citet{HeitjanRubin}, sampling univariate
$y$ values, used an acceptance-rejection procedure to draw quantities
analogous to our $W$ and $G$ from a confined bivariate normal
distribution. In our model, the correlation within $W_i$ and $G_i$
vectors poses a challenge to simulation. Note, however, that given the
subject-specific effects $b_i$ and $u_i$, the components of $W_i$ and
$G_i$ are independent. Thus, we can readily simulate $(W_i,G_i)$ from
the joint posterior of $(W_i,G_i,b_i,u_i)$. For each simulated $\theta$
and the observed data $y_i$, the posterior distribution of
$(W_i,G_i,b_i,u_i)$ is
\[
f(w_i,g_i,b_i,u_i|y_i,\theta) = f(w_i,g_i,b_i,u_i|\theta)
\frac{f(y_i|w_i,g_i,b_i,u_i,\theta)}{f(y_i|\theta)}.
\]
Because the values of $w_{it}$ and $g_{it}$ together determine
$y_{it}$, we have that
\[
f(y_i|w_i, g_i, b_i, u_i,\theta) =\prod_{t=1}^{m_i}I\bigl((w_{it},
g_{it})\in\mathrm{WG}(y_{it})\bigr),
\]
where $I$ is an indicator function. Accordingly,
\begin{eqnarray*}
&&
f(w_i, g_i, b_i, u_i|y_i, \theta)\\
&&\qquad\propto f(w_i, g_i, b_i, u_i|\theta)\prod_{t=1}^{m_i}I\bigl((w_{it},
g_{it})\in\mathrm{WG}(y_{it})\bigr) \\
&&\qquad = f(w_i, g_i|b_i, u_i, \theta)f(b_i,u_i|\theta)\prod
_{t=1}^{m_i}I\bigl((w_{it},g_{it})\in\mathrm{WG}(y_{it})\bigr) \\
&&\qquad = f(w_i|b_i,\theta)f(g_i|w_i, u_i,\theta)f(b_i,u_i|\theta)\prod
_{t=1}^{m_i}I\bigl((w_{it},g_{it})\in\mathrm{WG}(y_{it})\bigr) \\
&&\qquad = \Biggl(\prod_{t=1}^{m_i}f(w_{it}|b_i,\theta)f(g_{it}|w_{it},
u_i,\theta)I\bigl((w_{it},g_{it})\in\mathrm{WG}(y_{it})\bigr)\Biggr)\\
&&\qquad\quad{} \times f(b_i|\sigma_b)f(u_i|\sigma_u).
\end{eqnarray*}
Thus, given random effects $b_i$ and $u_i$, the imputation of
$(w_i,g_i)$ is obtained by independent draws of $(w_{it},g_{it})$,
$t=1,\ldots,m_i$, which can be implemented as an acceptance-rejection
procedure. We therefore impute the data as follows:
\begin{longlist}[(2)]
\item[(1)] Make independent draws, $\theta^{(k)}, k=1,\ldots,K$ from
$f(\theta|y_i)$ by SIR.
\item[(2)] Given $\theta^{(k)}$, for $i=1,\ldots,n$, independently draw
$b_i^{(k)}\sim N(0, \sigma_b^{(k)2})$ and $u_i^{(k)} \sim N(0,\sigma_u^{(k)2})$.
\item[(3)] For $i=1,\ldots,n$, given $\theta^{(k)}$ and $b_i^{(k)}$, for
$t=1,\ldots, m_i$, draw $w_{it}^{(k)}$ as Poisson with mean \eqref
{recall}. Then given $\theta^{(k)}$, $u_i^{(k)}$ and $w_{it}^{(k)}$,
draw misreporting type $g_{it}^{(k)}$ from \eqref{misreport}. If
$I((w_{it}^{(k)},g_{it}^{(k)})\in\mathrm{WG}(y_{it}))=0$, discard
$(w_{it}^{(k)},g_{it}^{(k)})$ and repeat this step until
$I((w_{it}^{(k)},g_{it}^{(k)})\in\mathrm{WG}(y_{it}))=1$.
\end{longlist}

To assess model fit, we plot $K$ histograms of the imputed latent count~$w$.
Implausible patterns in these histograms, such as peaks or troughs
at multiples of 5, suggest incorrect modeling of the heaping. We can
also base discrepancy diagnostics specifically on the fractions of
reported consumptions that are divisible by~5.

\section{Simulations}\label{sec4}
To examine the performance of our approach, we conducted simulations
replicating the structure of the Shiffman data with $m=12$ nonvisit-day
observations per subject. Each data set consisted of $n=100$ subjects,
and for simplicity we do not consider baseline covariates. For each
subject we first set $x_i$ as an observed EMA count vector from the
data and generated a random effect $b_i\sim N(0,\sigma_b^{2}=0.09)$. We
then generated $W_{it}$ values as independent Poisson deviates with
conditional mean \eqref{recall}. With $\beta_0 = 2.358$, $\beta_1 =
0.2628$, when $b_i = 0$ and EMA count $x_{it} = 20$, the mean latent
recall is 23.2, and when $x_{it} = 30$ it is~25.8. With the random
effect distributed as designated above, the marginal mean recalls for
$x_{it} = 20$ and $x_{it} = 30$ are 24.3 and 27.0, respectively.

Next we generated the latent heaping behavior indicator $G_{it}$ from
\eqref{misreport}. We set the parameters to their estimates from the
Shiffman data: the intercepts $\gamma_1$, $\gamma_2$, $\gamma_3$ were
$-1.485$, $-5.280$ and $-10.141$, respectively, and the slope $\gamma
_0$ was $0.1098$. We simulated the random effect $u_i \sim N(0,\sigma
_u^2=7.1)$. Under this setting, when $u_i=0$ and $w_{it}=22$, the
probability of exact reporting is 28.3\%, and the probabilities of
rounding to the nearest multiples of 5, 10 and 20 are 66.3\%, 5.4\% and
0.04\%, respectively. When the latent count $w_{it}=36$, these
probabilities are 7.8\%, 71.2\%, 20.8\% and 0.2\%, respectively. The
simulated latent $w_{it}$ and $g_{it}$ determined $y_{it}$ as
illustrated in Figure~\ref{heapdiag3}.

These parameter values allow for considerable discrepancy between
remembered and recorded consumption. To examine our methods when the
latent recall and EMA match more closely, we conducted a second
simulation under parameter values that gave better agreement. In this
scenario, we assumed $\beta_0=0$ and $\beta_1=1$ with $b_i \sim N(0,
0.05)$. Thus, when $b_i=0$, the expected precise recall ${\mathrm
E}(w_{it})=x_{it}$, and the marginal mean recalls are 20.5 and 30.8 for
EMA counts of 20 and 30, respectively. We set the parameters in the
heaping behavior models at $-1.07$, $-4.37$, $-6.52$ and $0.088$ for
$\gamma_1$, $\gamma_2$, $\gamma_3$ and $\gamma_0$, respectively, and
$\sigma^2_u = 5.9$. In this case, when $u_{it}=0$, the probabilities of
reporting exactly and to the nearest multiples of 5, 10 and 20 for a
true count of 22 are 29.6\%, 62.3\%, 7.1\% and 1\%, respectively.

%
\begin{table}
\caption{Results of 100 simulations of the
mis-remembering/heaping model}\label{simulation}
\begin{tabular*}{\tablewidth}{@{\extracolsep{\fill}}ld{3.3}d{2.3}cd{2.3}cc@{}}
\hline
&\multicolumn{1}{c}{\textbf{True}}&\multicolumn{1}{c}{\textbf{Mean of}} &
\multicolumn{1}{c}{\textbf{SD of}} &&& \multicolumn{1}{c@{}}{\textbf{Coverage of}}\\
\textbf{Parameter}&\multicolumn{1}{c}{\textbf{value}}
&\multicolumn{1}{c}{\textbf{estimate}}&\multicolumn{1}{c}{\textbf{estimate}}
&\multicolumn{1}{c}{\textbf{Bias}}
&\multicolumn{1}{c}{$\bolds{\sqrt{\mathrm{MSE}}}$}
&\multicolumn{1}{c@{}}{\textbf{95\% CI (\%)}}\\
\hline
\multicolumn{7}{@{}c@{}}{\textit{Case} 1: \textit{Estimated mis-remembering}}\\[4pt]
Latent recall\\
\quad$\beta_0$&2.36&2.36&0.07&0.002&0.07&95\\
\quad$\beta_1$&0.26&0.26&0.02&0.001&0.02&93\\
\quad$\sigma_b$&0.30&0.30&0.02&0.001&0.02&95
\\[4pt]
Heaping behavior\\
\quad$\gamma_1$&-1.49&-1.53&0.56&-0.04&0.56&94\\
\quad$\gamma_2$&-5.28&-5.31&0.66&-0.03&0.66&98\\
\quad$\gamma_3$&-10.14&-9.99&2.55&0.15&2.54&80\\
\quad$\gamma_0$&0.11&0.11&0.02&0.002&0.02&96\\
\quad$\sigma_u$&2.67&2.61&0.29&-0.06&0.29&98
\\[4pt]
\multicolumn{7}{@{}c@{}}{\textit{Case} 2: \textit{Minimal mis-remembering}}\\[4pt]
Latent recall\\
\quad$\beta_0$&0.0&-0.01&0.09&-0.01&0.09&94\\
\quad$\beta_1$&1.0&1.00&0.03&0.005&0.03&94\\
\quad$\sigma_b$&0.22&0.22&0.02&-0.001&0.02&97
\\[4pt]
Heaping behavior\\
\quad$\gamma_1$&-1.07&-1.08&0.43&-0.007&0.43&98\\
\quad$\gamma_2$&-4.37&-4.36&0.60&0.007&0.59&94\\
\quad$\gamma_3$&-6.52&-6.43&0.66&0.09&0.67&94\\
\quad$\gamma_0$&0.088&0.090&0.02&0.002&0.02&95\\
\quad$\sigma_u$&2.44&2.41&0.27&-0.02&0.27&95\\
\hline
\end{tabular*}
\end{table}

Table~\ref{simulation} presents summaries of 100 simulations of
estimates of the parameter $\theta=(\beta_0,\beta_1,\sigma_b,\gamma
_1,\gamma_2,\gamma_3,\gamma_0,\sigma_u)$. Under both scenarios, the
MLEs of the fixed-effect coefficients fell near the true values on
average, with no more than 0.5\% bias for the parameters in the recall
model and no more than 2.7\% bias for those in the heaping model. The
random effects variance estimates are also well estimated, with bias
less than 1\%. The coverage probabilities of nominal 95\% confidence
%
%
\begin{table}
\caption{Results of 100 simulations of the
mis-remembering/heaping model with parameters estimated from the data
(case 1) and SEs computed by the parametric bootstrap}\label{simulation2}
\begin{tabular*}{\tablewidth}{@{\extracolsep{\fill}}ld{3.2}d{3.2}cd{2.3}cc@{}}
\hline
&\multicolumn{1}{c}{\textbf{True}}
& \multicolumn{1}{c}{\textbf{Mean of}} &\multicolumn{1}{c}{\textbf{SD
of}}
&&&\multicolumn{1}{c@{}}{\textbf{Coverage of}}\\
\textbf{Parameter}&\multicolumn{1}{c}{\textbf{value}}
&\multicolumn{1}{c}{\textbf{estimate}}
&\multicolumn{1}{c}{\textbf{estimate}}
&\multicolumn{1}{c}{\textbf{Bias}}
&\multicolumn{1}{c}{$\bolds{\sqrt{\mathrm{MSE}}}$}
&\multicolumn{1}{c@{}}{\textbf{95\% CI (\%)}}\\
\hline
Latent recall\\
\quad$\beta_0$&2.36&2.36&0.08&-0.003&0.08&90\\
\quad$\beta_1$&0.26&0.26&0.02&0.001&0.02&90\\
\quad$\sigma_b$&0.30&0.30&0.02&-0.001&0.02&95
\\[4pt]
Heaping behavior\\
\quad$\gamma_1$&-1.49&-1.61&0.55&-0.12&0.56&94\\
\quad$\gamma_2$&-5.28&-5.42&0.69&-0.14&0.70&96\\
\quad$\gamma_3$&-10.14&-10.61&3.56&-0.47&3.58&87\\
\quad$\gamma_0$&0.11&0.11&0.02&0.005&0.02&95\\
\quad$\sigma_u$&2.67&2.64&0.32&-0.03&0.32&92\\
\hline
\end{tabular*}
\end{table}
intervals range from 93\% to 98\%, except for $\gamma_3$ in case 1,
where coverage is only 80\%. The poor coverage rate for this parameter
is a consequence of instability in the inverse Hessian matrix; it can
be improved by creating parametric bootstrap confidence intervals
(Table~\ref{simulation2}). The simulation shows good performance of the
MLEs, and, as the sample size is large, we expect the Bayesian
estimates to behave similarly. Moreover, the maximization part of the
MLE calculation can help identify multimodality of the likelihood,
should it occur, and singularity of the Hessian that we use in the
Bayesian sampling.\

\section{Data analysis}\label{sec5}
We applied the method of Section~\ref{sec2} to the Shiffman data, with
the aim of evaluating our posited two-stage process
as an explanation for the discrepancy between actual and reported
consumption. To focus on the link between the self-report and true
count, our first analysis included only log EMA count in \eqref{recall}
and a visit day indicator in \eqref{misreport}. The latter is important
because it seems reasonable that distance in time from the event would
be a strong predictor of heaping coarseness. Our second analysis
expanded the recall model to include a range of baseline
characteristics: demographics (age, sex, race and education);
addiction; measures of nicotine dependence [the Fagerstr\"{o}m Test for
Nicotine Dependence (FTND) and the Nicotine Dependence Syndrome Scale
(NDSS)]; and EMA compliance measured as the daily percentage of missed
prompts. Age, education, FTND and EMA compliance are considered as
quantitative variables, sex and race are binary indicators, and
addiction is a categorical variable taking three levels (possible,
probable and definite). They are the first variables that a smoking
researcher would think to investigate, and could potentially affect
remembered count or heaping probability. The two measures of nicotine
dependence FTND and NDSS showed only a modest correlation, with
Spearman $r = 0.56$ in our data. So we considered both in the model.
The data set and programming code are included in the supplementary
materials [\citet{data}].

\subsection{Evaluating goodness of fit}\label{sec5.1}
We evaluated model fit by creating multiple draws from the posterior
predictive distribution of latent quantities as discussed in Section
\ref{sec3}.
Lack of smoothness in the histogram of the imputed latent count would
suggest an inadequate heaping model.

%
\begin{figure}

\includegraphics{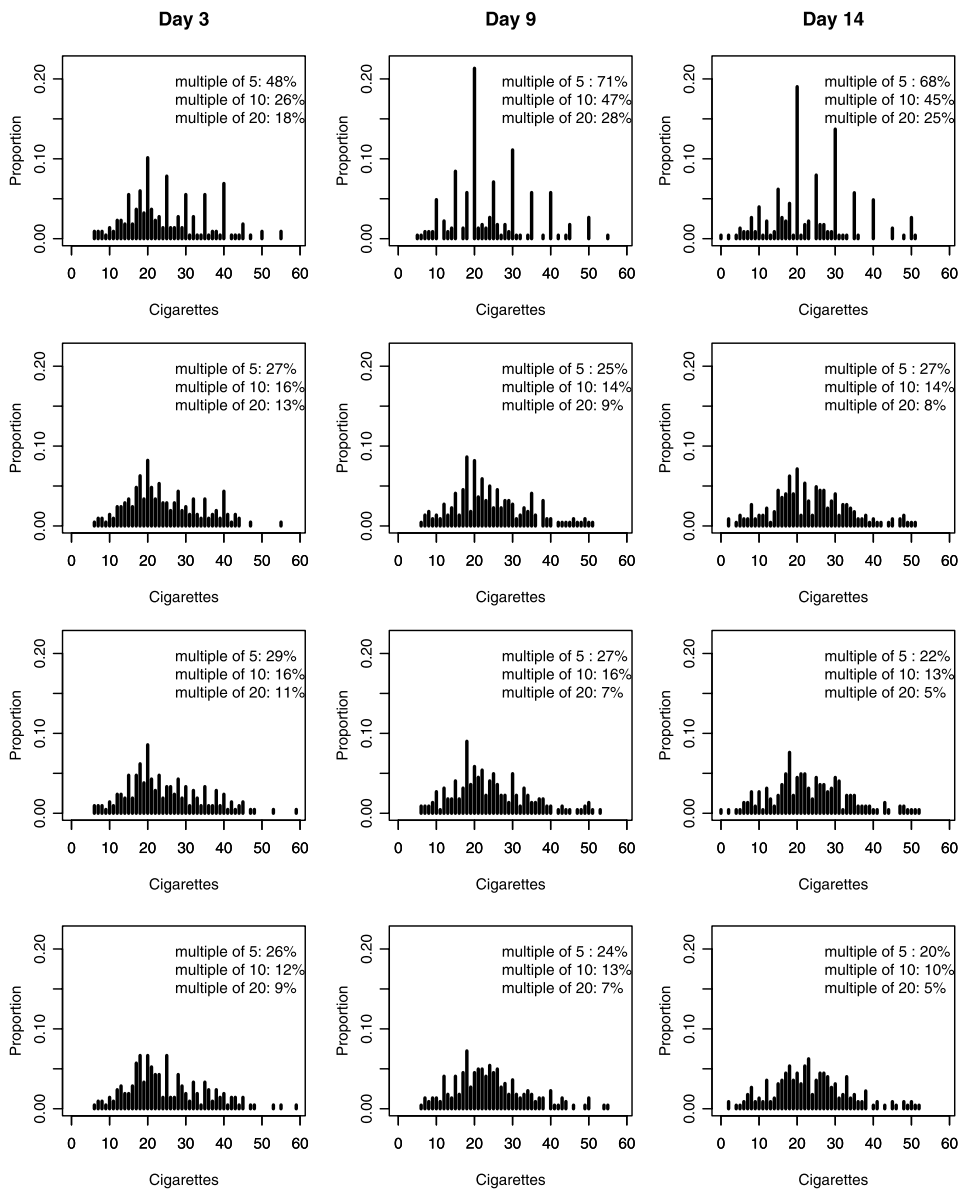}%
\vspace*{-2pt}
\caption{Top row: histogram of self-reported cigarette
consumption. Lower three rows: histograms of draws from the posterior
distribution of the latent exact consumption
recall.}\label{ppc}\vspace*{-4pt}
\end{figure}

We evaluated goodness of fit for the model that includes log EMA count
in \eqref{recall} and a visit day indicator in \eqref{misreport}. The
top row in Figure~\ref{ppc} displays the histograms of TLFB cigarette
consumption at days 3 (a visit day), 9 and 14. The spikes at 10, 15,
20, 25, 30, etc. are characteristic of self-reported cigarette counts
[\citet{WangHeitjan}]. As many as 70\% of subjects reported
cigarette smoking in multiples of 5 for nonvisit-day consumption,
whereas for the visit day (day 3) that number is only 48\%. Only $1/4$ of
the counts on the visit day ended in $0$.

The next three rows represent independent draws of the latent count
$W_{it}$. The spikes at multiples of 20, 10 or 5 have disappeared.
Compared to the self-reported count, the percentage of subjects whose
exact counts are divisible by 5 (or 10 or~20) is smaller and consistent
across time. Averaged over three imputations, the fraction of counts
ending in multiples of $5$ is 27\%, 25\% and 23\% on days 3, 9 and 14,
respectively, and 15\%, 14\% and 12\% end in multiples of $10$. These
checks indicate that our model offers a plausible explanation for the heaping.

\subsection{The fitted model}\label{sec5.2}
In order to assess the impact of the assumed correlation structure, we
fit the model as proposed in \eqref{recall} and \eqref{misreport} and
also a model that excludes random effects. Posterior modes and 95\%
credible intervals (CIs) appear in Tables~\ref{posterior1} and \ref
{posterior2}. The estimates in both the remembered count model that
%
%
\begin{table}
\caption{Estimated parameters from the Shiffman data
under simple models for recall (EMA only) and heaping (remembered count
and visit day indicator)}\label{posterior1}
\begin{tabular*}{\tablewidth}{@{\extracolsep{\fill}}ld{3.2}cd{2.2}c@{}}
\hline
&\multicolumn{2}{c}{\textbf{Random effects model}} &\multicolumn
{2}{c@{}}{\textbf{Independence model}}\\[-4pt]
& \multicolumn{2}{c}{\hrulefill} & \multicolumn{2}{c@{}}{\hrulefill}\\
&\multicolumn{1}{c}{\textbf{Posterior}}& &\multicolumn{1}{c}{\textbf{Posterior}}&\\
\textbf{Parameter}&\multicolumn{1}{c}{\textbf{mode}}&\multicolumn{1}{c}{\textbf{95\% CI}}
&\multicolumn{1}{c}{\textbf{mode}}&\multicolumn{1}{c@{}}{\textbf{95\% CI}}\\
\hline
\multicolumn{5}{@{}l}{Latent recall: Poisson model}\\
\quad Intercept: $\beta_0$&2.32&$[2.24, 2.40]$ &
1.14&$[1.09, 1.20]$\\
\quad$\ln$(EMA): $\beta_1$&0.27&$[0.25, 0.30]$ &
0.68&$[0.66, 0.69]$\\
\quad$\sigma_b^2$&0.09&$[0.08, 0.11]$&\\
[4pt]
\multicolumn{5}{@{}l}{Heaping behavior: proportional odds model}\\
\quad Intercept 1: $\gamma_1$&-1.50&$[-2.17, -0.85]$
&-1.06&$[-1.30, -0.84]$\\
\quad Intercept 2: $\gamma_2$&-5.21&$[-6.14, -4.43]$
&-2.94&$[-3.26, -2.65]$\\
\quad Intercept 3: $\gamma_3$&-10.15&$[-12.49, -8.48]$
&-4.17&$[-4.59, -3.82]$\\
\quad Exact count (latent): $w$&0.11&$[0.09, 0.13]$&
0.07&$[0.06, 0.08]$\\
\quad Visit day&-2.96&$[-3.50, -2.50]$ &
-1.29&$[-1.54, -1.06]$\\
\quad$\sigma_u^2$&6.65&$[5.12, 9.08]$ &\\
\hline
\end{tabular*}
\end{table}
characterizes the latent recall process and the heaping behavior model
are sensitive to the assumption of random effects. The Bayesian
information criterion (BIC) of the model with two random effects is
14,705 when including EMA as the only predictor and 14,059 when
including EMA and the baseline patient characteristic predictors. The
BICs for the corresponding models excluding random effects are 18,340
and 16,641, respectively.\vadjust{\goodbreak} Thus, the evidence is overwhelming that the
mixed model is preferable. Furthermore, we included the patient
characteristic predictors as covariates in both the remembered count
model and heaping process model, but this model ($\mbox{BIC} = 14\mbox{,}079$) is
less favorable compared to the model with the covariates in just the
latent remembered count model. None of these predictors is significant
in the heaping process model (results not shown).\vadjust{\goodbreak}

%
\begin{table}
\caption{Estimated parameters from the Shiffman data
under an expanded model for recall}\label{posterior2}
\begin{tabular*}{\tablewidth}{@{\extracolsep{\fill}}ld{3.3}cd{2.3}c@{}}
\hline
&\multicolumn{2}{c}{\textbf{Random effects model}} &\multicolumn
{2}{c@{}}{\textbf{Independence model}}\\[-4pt]
& \multicolumn{2}{c}{\hrulefill} & \multicolumn{2}{c@{}}{\hrulefill}\\
&\multicolumn{1}{c}{\textbf{Posterior}}& &\multicolumn{1}{c}{\textbf{Posterior}}&\\
\textbf{Parameter}&\multicolumn{1}{c}{\textbf{mode}}&\multicolumn{1}{c}{\textbf{95\% CI}}
&\multicolumn{1}{c}{\textbf{mode}}&\multicolumn{1}{c@{}}{\textbf{95\% CI}}\\
\hline
\multicolumn{5}{@{}l}{Latent recall: Poisson model}\\
\quad Intercept: $\beta_0$&2.34&$[2.21, 2.49]$ &
1.51&$[1.44, 1.58]$\\
\quad$\ln$(EMA): $\beta_1$&0.25&$[0.23, 0.28]$ &
0.53&$[0.51, 0.55]$\\
\quad Addicted \\
\qquad Possible vs. definite&0.07&$[-0.10, 0.24]$
&0.05&$[0.01, 0.09]$\\
\qquad Probable vs. definite&- 0.01&$[-0.11, 0.08]$
&-0.02&$[-0.04, 0.006]$\\
\quad FTND &0.06&$[0.04, 0.08]$ &0.04&$[0.03, 0.05]$\\
\quad NDSS &0.08&$[0.05, 0.12]$ &0.05&$[0.04, 0.06]$\\
\quad EMA compliance &0.13&$[-0.28, 0.51]$ &
0.39&$[0.29, 0.49]$\\
\quad Age &0.002&$[-0.001, 0.006]$ &0.003&$[0.002,
0.004]$\\
\quad Race (black vs. white) &-0.14&$[-0.27, -0.01]$
&-0.06&$[-0.10, -0.03]$\\
\quad Sex (male vs. female) &0.16&$[0.10, 0.23]$&
0.12&$[0.09, 0.23]$\\
\quad Education &-0.001&$[-0.03, 0.02]$ &
0.003&$[-0.004, 0.009]$\\
\quad$\sigma_b^2$&0.06&$[0.05, 0.07]$&\\
[4pt]
\multicolumn{5}{@{}l}{Heaping behavior: proportional odds model}\\
\quad Intercept 1: $\gamma_1$&-1.62&$[-2.35, -0.90]$
&-1.14&$[-1.37, -0.91]$\\
\quad Intercept 2: $\gamma_2$&-5.52&$[-6.42, -4.61]$
&-3.15&$[-3.47, -2.82]$\\
\quad Intercept 3: $\gamma_3$&-10.31&$[-12.65, -8.37]$&
-4.54&$[-4.99, -4.08]$\\
\quad Exact count: $w$&0.11&$[0.09, 0.14]$ &
0.07&$[0.06, 0.08]$\\
\quad Visit day&-2.99&$[-3.51, -2.47]$ &
-1.26&$[-1.50, -1.02]$\\
\quad$\sigma_u^2$&6.79&$[4.73, 8.68]$ &\\
\hline
\end{tabular*}
\end{table}

The 95\% CI of $\beta_1$ is $[0.23, 0.28]$, indicating that remembered
consumption is positively associated with recorded EMA consumption. In
addition, baseline patient characteristics FTND, NDSS, race and gender
have significant effects on the recall process. For fixed EMA count,
the following characteristics are associated with greater remembered
smoking: higher nicotine dependence (measured by both FTND and NDSS),
white ethnicity (compared to black) and male sex.

\begin{figure}

\includegraphics{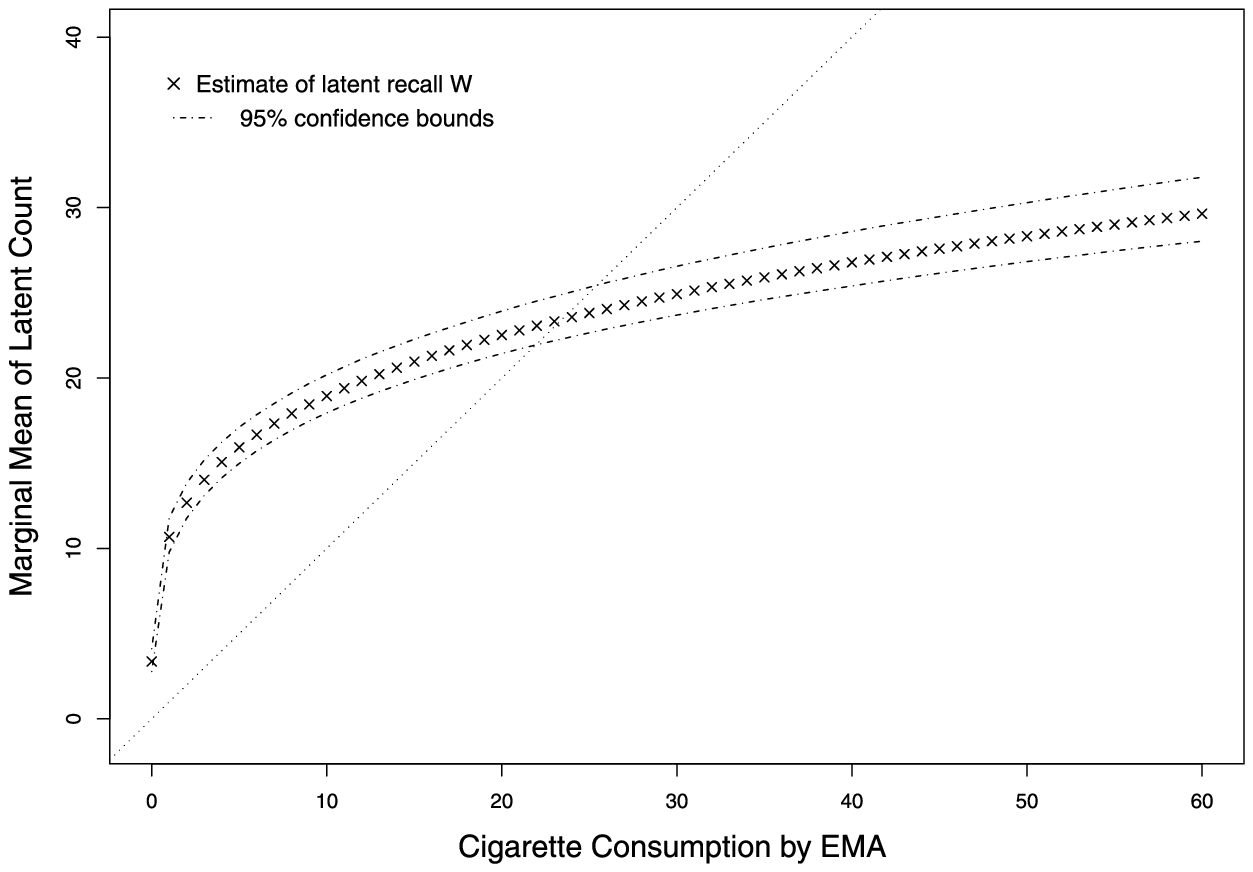}

\caption{Estimate of the conditional mean of recalled
count given EMA count in the Poisson mis-remembering model. Covariates
are fixed at education${}={}$high school, addicted${}={}$definitely, race${}={}$white,
sex${}={}$female, and mean values of the quantitative predictors:
FTND${}={}$5.97, $\mbox{NDSS}=-0.023$, age${}={}$43.5,
and EMA noncompliance${}={}$10.1\%.}
\label{expw}
\end{figure}

Figure~\ref{expw} displays the estimated curve of the mean of $W_{it}$
against the EMA count. A natural hypothesis is that the estimated
latent mean agrees with EMA, which would be reflected in the Poisson
model by an estimated intercept of $0$ and slope of~$1$; one might call
this a model of unbiased memory. To the contrary, Figure~\ref{expw}
shows that the fitted mean curve diverges substantially from the
$45^{\circ}$ line, with the lighter smokers on average overestimating
their consumption and the heavier smokers underestimating consumption.
The mean remembered consumption agrees with the true count roughly in
the range 22--26 cigarettes, or slightly more than a pack per
day.

%

\begin{figure}

\includegraphics{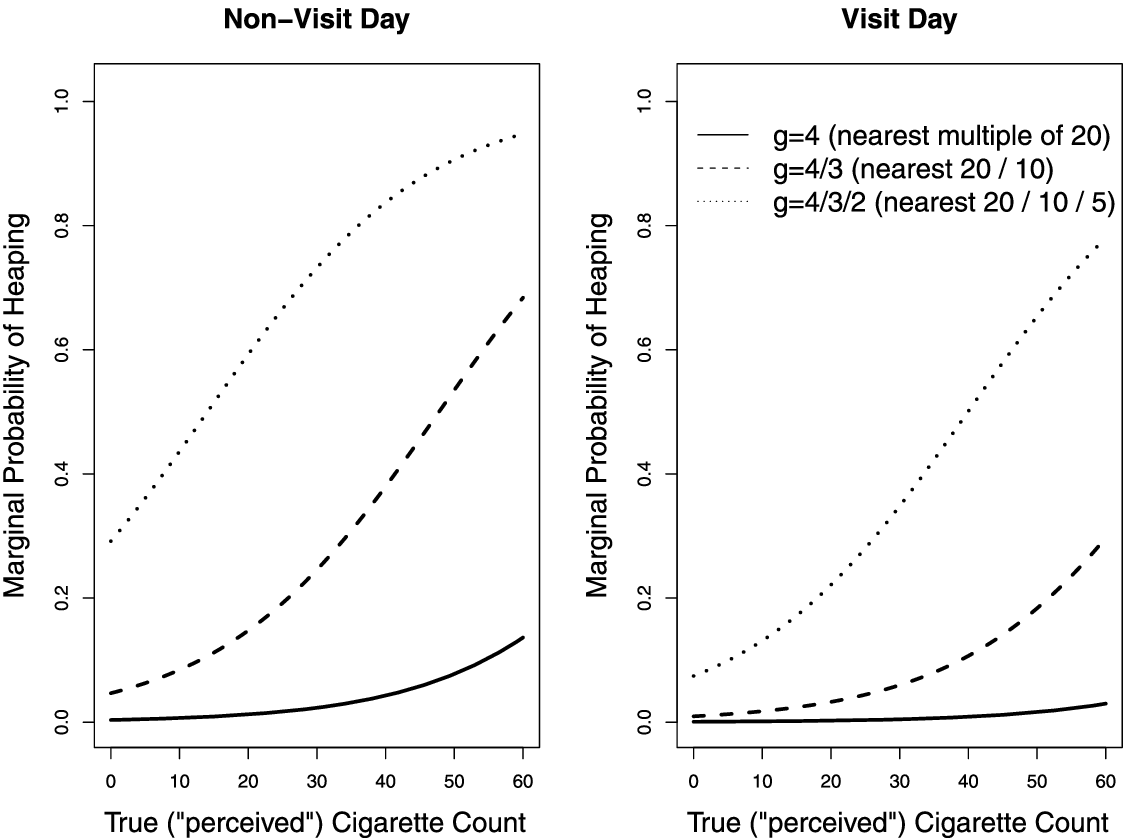}

\caption{Estimated rounding behavior given EMA count
in the proportional odds heaping model.}\label{heapprob}
\end{figure}

Figure~\ref{heapprob} shows the estimated heaping probability as a
function of remembered cigarette consumption for visit and nonvisit
days. The possibility of rounded-off reporting increases rapidly as the
remembered count increases, although surprisingly the probability of
rounding to the nearest 20 is not large for either type of day. When
the perception of smoking is more than two packs, say, 41 cigarettes,
the chance of heaped reporting rises to more than 84\%, of which 37\%
is attributed to half-pack rounding. The results confirm that the
degree of heaping is much smaller on visit days. For example, only 51\%
of subjects round off the visit-day count when reporting 41 cigarettes,
and among those 39\% round off to the nearest multiple of 5.

\section{Discussion}\label{sec6}
We have developed a model to describe the process whereby exact
longitudinal measurements become distorted by retrospective recall. Our
approach uses latent processes to explain the data as a result of
mis-remembering and rounding: a model of the latent exact value
describes subject-level recall and allows for association over time and
with baseline predictors, while a misreporting model describes the
dependence of heaping coarseness on the latent value and other
predictors. Random effects represent individual propensities in recall
and heaping; in our data, inferences depend strongly on the inclusion
of these random effects.

%

The data suggest that both mis-remembering and heaping contribute
substantially to the distortion of cigarette counts. The curve of mean
remembered count as a function of EMA count departs markedly from the
$45^{\circ}$ line, with lighter smokers overstating consumption and
heavier smokers understating consumption. The remembered smoking
coincides with the accurate EMA count at around 24~cigarettes,
suggesting that the popularity of reporting one pack per day is
partially a result of the general heaping behavior rather than a
particular affinity for remembering a pack a day. The curves of heaping
probabilities suggest that exact reporting is uncommon and practically
disappears beyond about 40 cigarettes/day. Nevertheless, it is
interesting just how much of the misreporting is due to
mis-remembering. The remembered cigarette consumption depends not only
on true consumption, but also on the subject's sex, race and degree of
nicotine dependence.

The interpretation of our model components as representing memory and
rounding depends on the assumption that EMA data are exact. Of course,
even EMA data are subject to errors, as smokers may neglect to record
cigarettes both at the time of smoking and later. Yet good
correspondence with smoking biomarkers strongly supports the use of EMA
over TLFB as a proxy for the truth [\citet{Shiffman2}].

We have implemented our model with a combination of standard numerical
methods including Gaussian quadrature, quasi-Newton
optimization and sampling-importance resampling. Our experience
suggests that with the model as specified, and incorporating a modest
numbers of predictors, the method is robust and efficient. Increasing
the number of random effects would increase the time demands (from the
numerical integration) and raise the possibility of numerical
instability (from possible errors in integration). For more extensive
models, sophisticated approaches based on MCMC sampling would be
necessary.

Our model allows for the inclusion of covariates to better explain the
discrepancy between smokers' self-perceived behaviors and reality. It
also provides a basis for predicting true counts (effectively the EMA
data) from reported TLFB counts. This would be a valuable activity in
the large number of studies that do not collect EMA data. To predict
true counts from the recalled counts, we first need to estimate the
parameters $\theta$ in the model using a subset of the primary study or
an external independent study that collects both TLFB count $Y$ and
accurate EMA count $X$. Then we can impute the true count together with
the latent remembered count and heaped reporting behavior.
Specifically, the posterior distribution of $(W_i,G_i,x_i,b_i,u_i)$ is
\begin{eqnarray*}
&&f(w_i,g_i,x_i, b_i,u_i|y_i,\theta)\\
&&\qquad = f(w_i,g_i,x_i, b_i,u_i|\theta)
\frac{f(y_i|w_i,g_i,x_i, b_i,u_i,\theta)}{f(y_i|\theta)}\\
&&\qquad \propto \Biggl(\prod_{t=1}^{m_i}f(w_{it}|x_{it},b_i,\theta
)f(g_{it}|w_{it}, u_i,\theta)I\bigl((w_{it},g_{it})\in\mathrm{WG}(y_{it})\bigr)\Biggr)\\
&&\qquad\quad{} \times f(x_i)f(b_i|\sigma_b)f(u_i|\sigma_u),
\end{eqnarray*}
where $f(x_i)$ is the density function of the true count. Imputation
follows similar steps as described in Section~\ref{modelchecking} with
$\theta$ set equal to the maximum likelihood estimates.

The methods developed here also can have application in a wide variety
of settings in social and medical science involving self-reported
data---for example, assessing sexual risk behavior, trial drug
consumption, eating episodes and financial expenditures.

\section*{Acknowledgments}

We are grateful to two Associate Editors and a referee, whose
perceptive comments and suggestions greatly improved the paper.

\begin{supplement}
\stitle{Data and programming code for the analysis\\}
\slink[doi,text={10.1214/12-\break AOAS557SUPP}]{10.1214/12-AOAS557SUPP} 
\slink[url]{http://lib.stat.cmu.edu/aoas/557/dataandcode.zip}
\sdatatype{.zip}
\sdescription{It contains the daily TLFB and EMA data set, and SAS
and R code to implement the method.}
\end{supplement}

%

\printaddresses

\end{document}